\documentstyle[prc,aps,epsfig,twocolumn]{revtex}
\begin{document}

\baselineskip=1. \baselineskip
\draft
\title{Self-consistent solution of Galitskii-Feynman equations at finite 
temperature}

\author{     P. Bo\.{z}ek~\footnote[1]{electronic
address~:~bozek@solaris.ifj.edu.pl}  }
\address{ NSCL, Michigan State University, East Lansing, MI-48824 \\  and \\
 Institute of Nuclear Physics, PL-31-342 Krak\'{o}w, Poland}
\date{\today}
\maketitle 

\begin{abstract}
We solve the in-medium T-matrix equation at finite temperature including the 
off-shell propagation of nucleons. In this way a self-consistent 
spectral function for the nucleons is obtained. The results are compared 
to a calculation using the quasiparticle approximation 
in the T-matrix equation. Also the effective in-medium cross sections for 
the two cases are compared.
\end{abstract}


\narrowtext

\section{Introduction}

The calculation of properties of a strongly interacting many-body 
system, is a challenging problem. 
Many efforts have been 
devoted to  calculations of the nuclear matter properties, both 
at zero and at
finite temperature. Most of those studies have been restricted to the
quasiparticle approximation. The quasiparticle approximation can 
be justified in the vicinity of the Fermi energy at zero temperature 
\cite{fw}. Explicit calculation show that the width of the  spectral function 
approaches zero when the energy approaches the Fermi energy and the 
temperature tends to zero \cite{baldo}.
However the knowledge of the whole spectral function is needed for a 
self-consistent Brueckner or Galitskii-Feynman calculation.
Moreover at larger temperatures there is no region in the momentum space 
where the Pauli blocking reduces the scattering width. If one wants to 
address the dynamics of  heavy ion collisions at energies of
 a few hundreds of MeV per nucleon the knowledge of effective cross sections
in  dense and excited nuclear nuclear matter is needed. The calculation of 
the in-medium T-matrix (or the Brueckner G-matrix)
 gives an estimate of this cross section \cite{cs}. 
The calculation of the in-medium cross section
(at equilibrium)
 can and should take into account the off-shell propagation of the nucleons.
The off-shell propagation of scattering 
 nucleons changes quantitatively the value of 
the cross section  and in the case of soft emission it changes also 
its qualitative behavior \cite{kn_vo,moj2}.

  The existing calculations 
of the nuclear spectral functions assume a quasiparticle approximations 
in the summation of the ladder diagrams \cite{nm,baldo,vo,ko,di1,alm,rop}.
From the knowledge of the T matrix (in the Galitskii-Feynman equations) or 
the G matrix (in the Brueckner equations) the imaginary part of the
one-particle self-energy can be calculated \cite{kb,paw1,bm}. 
However, it has been noted already several years ago, 
 that calculations of the self-energy in the Born approximation
using a semiclassical
collision term and a quantum collision term with off-shell propagators give 
different results \cite{paw2}. The semiclassical one-particle width
 being generally larger than the self-consistent quantum one.
Clearly the spectral functions obtained in  the quasiparticle approximation
should be checked in a self-consistent calculation with in-medium off-shell 
nucleon propagators. 

A spectral function obtained in a self-consistent 
way would provide very important 
information about nuclear matter and its behavior both at zero and
 at finite temperatures.
We could mention in this respect the  electron scattering on nuclei \cite{es},
the subthreshold particle production \cite{st,moj1}, the calculation of 
in-medium
 effective cross sections \cite{cs}, the backward scattering \cite{paw3}
and of course a self consistent calculation of the saturation energy and 
the properties of the nuclear matter \cite{nm,baldo,vo,ko,di1,rop}.
 In the last example
  the spectral function is needed not 
only for the off-shell propagation of nucleons in the ladder diagrams.
It enters also in the calculation of the Hartree-Fock energy (through the 
calculation of the momentum distribution in an interacting system) and 
the dispersive contribution to the real  part of the 
self-energy.
The need for a self-consistent calculation was of course 
recognized, but real calculation have not been performed, except for restricted
kinematical conditions \cite{piongas}. Only very recently 
the off-shell nucleon propagation and scattering were addressed \cite{di2}. 

In this work we present an exploratory self-consistent calculation of the 
nucleon spectral function at finite temperature in the T-matrix approximation.
In the nuclear matter the T-matrix approximation leads to pairing transition 
at low temperatures \cite{rop,alm,di1}. 
We do not intend to address here 
the superfluity transition at low temperatures. 
This means of course that we stay at temperatures above the pairing transition.
The formalism to treat the pairing in the T-matrix approximation is still
 missing both in the quasiparticle as well as in the self-consistent version
\cite{alm,rop}.
It should be pointed out that similar self-consistent calculations can be
also
performed in the Brueckner scheme at low temperatures where no pairing 
instability occurs.

\section{In-medium T-matrix}

In the present work we use the real-time Green's functions formalism 
\cite{kb,paw1,bm},
which we found very suitable for calculations at finite temperature in the
 Born approximation \cite{moj2}. 
In equilibrium the Green's functions are defined by the spectral function~:
\begin{eqnarray}
G^<(p,\omega)=i f(\omega) A(p,\omega) \nonumber \\
G^>(p,\omega)=-i\big(1-f(\omega)\big) A(p,\omega) \ ,
\end{eqnarray}
where 
\begin{equation}
f(\omega)=\frac{1}{e^{(\omega-\mu)/T}+1}
\end{equation}
is the Fermi distribution and the spectral function $A$ is 
\begin{equation}
A(p,\omega)=-2{\rm Im} G^+(p,\omega) \ .
\end{equation}
$G^\pm$ denote the retarded(advanced) Green's function.
The spectral function can be written equivalently using the self-energy
\begin{equation}
\label{spfu}
A(p,\omega)=\frac{\Gamma(p,\omega)}{\big(
(\omega-p^2/2 m-{\rm Re} \Sigma^+(p,\omega))^2+\Gamma(p,\omega)^2/4 \big)} \ ,
\end{equation}
where
\begin{equation}
\Gamma(p,\omega)=-2{\rm Im}\Sigma^+(p,\omega) \ .
\end{equation}
In order to reach a consistent approximation scheme we have to calculate the
retarded self-energy $\Sigma^+$ using the one-particle Green's functions $G$.
In a previous work we have calculated the self-energy in the Born approximation
\cite{moj2}. Below we address the calculation in the more 
complicated T-matrix approximation \cite{kb,paw1,bm}. This approximation
takes into account  the  two-body correlations and thus becomes exact in a 
dilute system, but is not restricted to zero or low temperatures as
 the Brueckner approximation. The nuclear matter at normal nuclear density is 
not a dilute interacting system and correction from three body correlation 
are probably non-negligible. However, the T-matrix ladder resummation in-medium
 represents a serious improvement over the Born approximation and gives good
approximations for  one-particle properties of the system.

\begin{figure}
\begin{center}
\epsfig{file=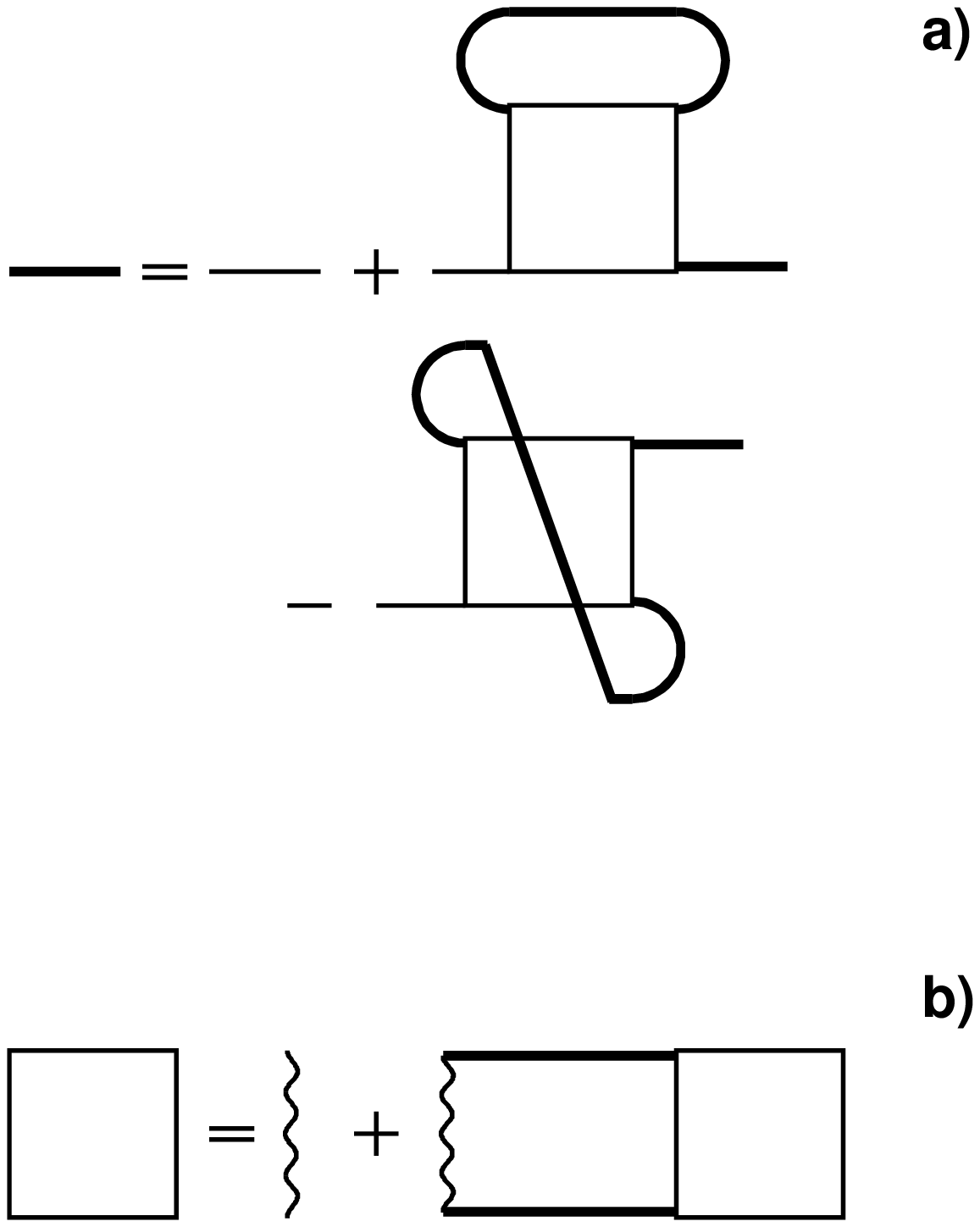,width=0.5\textwidth}
\end{center}
\caption{{\bf a)} Diagrammatic representation  
of the self-energy in the T-matrix approximation. 
The self-energy represented in the  diagram a) 
leads to Eqs. (\ref{ims}) and (\ref{resi}). \\
{\bf b)} Diagrammatic representation of the T-matrix equation (\ref{teq}).
 The thin lines represent
the noninteracting  fermion propagators, the thick lines  are interacting
 off-shell propagators and
the wavy lines denote the interaction potential. In the quasiparticle 
approximation the in-medium off-shell propagator is replaced by the 
on-shell propagator in the intermediate states in the T-matrix equation
and in the loops of the self-energy diagrams.}
\label{tmatfig}
\end{figure}

The T-matrix (Fig. \ref{tmatfig})  for a system with a
two-body interaction $V({\bf p},{\bf p}^{'})$ is defined as \cite{kb,paw1,bm}~:
\begin{eqnarray}
\label{teq}
<{\bf p}|T^\pm({\bf P},\omega)|{\bf p}^{'}> = V({\bf p},{\bf p}^{'})+ 
 \int\frac{d^3k}{(2 \pi)^3}
\int\frac{d^3q}{(2 \pi)^3} \nonumber \\ V({\bf p},{\bf k}) 
<{\bf k}|{\cal G}^\pm({\bf P},\omega)|{\bf q}>
 <{\bf q}|T^\pm({\bf P},\omega)
|{\bf p}^{'}> \ ,
\end{eqnarray}
where the disconnected two-particle propagator is~:
\begin{eqnarray}
\label{twpro}
<{\bf p}|{\cal G}^\pm({\bf P},\Omega)|{\bf p}^{'}> =  
(2 \pi)^3 \delta^3({\bf p}-{\bf p}^{'})\int \frac{d\omega^{'}}{2 \pi}
\int \frac{d\omega}{2 \pi} \nonumber \\
\bigg( G^<({\bf P}/2+{\bf p},\omega-\omega^{'})G^<({\bf P}/2-{\bf p},
\omega^{'}) \nonumber \\ -G^>({\bf P}/2+{\bf p},\omega-
\omega^{'})G^>({\bf P}/2-{\bf p},\omega^{'}) \bigg) \nonumber \\
/ \  \bigg(\Omega -\omega \pm 
i\epsilon \bigg) \ .
\end{eqnarray}
Taking in the above expression only the particle-particle propagator 
(the $G^>G^>$ factor)
results in the Brueckner approximation. At this point the quasiparticle 
approximation is usually made
\begin{equation}
A(p,\omega)=2\pi Z \delta(\omega-\omega_p) \ ,
\end{equation}
where the single-particle energy is the solution of
\begin{equation}
\omega_p=\frac{p^2}{2m}+{\rm Re}\Sigma^+(p,\omega_p) \ 
\end{equation}
and 
\begin{equation}
Z^{-1}
=\Big(1-\frac{\partial {\rm Re}\Sigma^+(p,\omega)}
{\partial \omega}\Big)_{\omega=\omega_p}
\ .
\end{equation}
Performing the calculations in
 the framework of the quasiparticle approximation we 
shall put however $Z=1$ as has been done in many works \cite{alm,ko,vo}.
This results in the on-shell two particle propagator
\begin{eqnarray}
\label{twproqp}
<{\bf k}|{\cal G}^\pm({\bf P},\Omega)|k^{'}> =  
(2 \pi)^3 \delta^3({\bf k}-{\bf k}^{'}) \nonumber \\
\frac{1-f(\omega_{p_1})-f(\omega_{p_2}) }{\Omega -\omega_{p_1}-
\omega_{p_2} \pm i\epsilon} \  ,
\end{eqnarray}
with ${\bf p}_{1,2}={\bf P}/2\pm {\bf k}$.
Within the quasiparticle approximation the T-matrix equation takes the 
familiar form
\begin{eqnarray}
\label{teqqp}
<{\bf p}|T^\pm({\bf P},\omega)|{\bf p}^{'}>= V({\bf p},{\bf p}^{'}) + 
\int\frac{d^3k}{(2 \pi)^3} V({\bf p},{\bf k}) \nonumber \\
 \frac{1-f(\omega_{p_1})-f(\omega_{p_2})}{\omega
-\omega_{p_1}-\omega_{p_2}\pm i \epsilon} 
 <{\bf k}|T^\pm({\bf P},\omega)
|{\bf p}^{'}> \ .
\end{eqnarray}
We are using an angular averaged two particle propagator ${\cal G}^\pm$
both in the the self-consistent equation (\ref{teq}) and in its quasiparticle 
counterpart (\ref{teqqp}). This standard approximation for in-medium 
calculation allows to perform a partial wave expansion of the T-matrix.

The off-shell propagation of nucleons means that the spectral function is 
not sharply peaked around $\omega_p$, also it cannot be approximated by
putting a frequency independent width $\Gamma(p)$ in  Eq. 
(\ref{spfu}). This implies that the two frequency integrals in (\ref{twpro})
have to be done numerically. The two particle off-shell 
propagator ${\cal G}^\pm$
can then be used to calculate the T-matrix using  (\ref{teq}).
The equation for the T-matrix is an integral equation, but  in the
 present work we use 
a separable potential and the solution of Eq. (\ref{teq}) is 
trivial. However we make no simplifying assumptions concerning the 
one-particle spectral functions, so that the intermediate 
two particle propagator ${\cal G}^\pm$
 takes into account the off-shell nucleon propagation.

The imaginary part of the self-energy can on its turn be obtained 
from the T-matrix
\begin{eqnarray}
\label{ims}
{\rm Im}
\Sigma^+(p,\Omega) =\int\frac{d\omega}{2 \pi}\int \frac{d^3k}{(2 \pi)^3}
A(k,\omega) \nonumber \\
 <({\bf p}-{\bf k})/2|{\rm Im}T^+({\bf p}
+{\bf k},\Omega+\omega)|({\bf p}-{\bf k})/2>_A
\nonumber \\
 \Big( f(\omega)+g(\omega+\Omega) \Big) \ ,
\end{eqnarray}
where the index $A$  indicates that the T-matrix is antisymmetrized and 
\begin{equation}
g(\omega)=\frac{1}{e^{(\omega-2\mu)/T}-1} 
\end{equation}
is the Bose distribution. The relation (\ref{ims}) is true at equilibrium.
The explicit expressions for the self-energies in the partial wave 
expansion of the T-matrix can be found e.g. in \cite{BU}.
Again in the quasiparticle approximation Eq. (\ref{ims})
simplifies to
\begin{eqnarray}
\label{imsqp}
{\rm Im}\Sigma^+(p,\Omega) =\int \frac{d^3k}{(2 \pi)^3}
\Big( f(\omega_k)+g(\omega_k+\Omega) \Big) 
\nonumber \\
 <({\bf p}-{\bf k})/2|{\rm Im}T^+({\bf p}+{\bf k},\Omega+\omega_k)|(
{\bf p}-{\bf k})/2>_A \ .
\end{eqnarray}
The real part of the self-energy can be obtained using the dispersion relation
\begin{equation}
\label{resi}
{\rm Re}\Sigma(p,\omega)=\Sigma_{HF}(p)+{\cal P}\int\frac{d\omega^{'}}{2 \pi}
\frac{\Gamma(p,\omega^{'})}{\omega-\omega^{'}} \ ,
\end{equation}
where the Hartree-Fock energy is given by
\begin{eqnarray}
\label{hfeq}
\Sigma_{HF}(p)=-i\int\frac{d\omega^{}}{2 \pi}\int  \frac{d^3k}{(2 \pi)^3}
V(({\bf p}-{\bf k})/2,({\bf p}-{\bf k})/2)_A 
\nonumber \\ G^<(k,\omega) 
\end{eqnarray}
and in the quasiparticle approximation it is
\begin{equation}
\label{hfeqqp}
\Sigma_{HF}(p)=\int  \frac{d^3k}{(2 \pi)^3}
V((p-k)/2,(p-k)/2)_A f(\omega_k) \ .
\end{equation}

The set of equations (\ref{spfu}),
(\ref{teq}), (\ref{ims}) and (\ref{resi}) must be solved self-consistently.
The numerical solution  of this set of equations is done by iteration starting
 with a spectral function with finite, constant width and with the Hartree-Fock
part of ${\rm Re}\Sigma^+$. Then the T-matrix is calculated from Eq.
(\ref{teq}) and the result is used to obtain the imaginary part of the
self-energy from 
(\ref{ims}). The iterations are performed until the self-energy becomes stable.
 Typically around 10 iterations are needed if the starting values of the 
Fermi energy and the initial value of the single-particle  width are
of the right order of magnitude.

The quasiparticle approximation in this work means that the T-matrix
is calculated from the formula (\ref{teqqp}),
where the single-particle energies are determined by the Hartree-Fock energy 
 (\ref{hfeqqp}). The T-matrix in the quasiparticle approximation is then used 
to calculate the  imaginary part of the self-energy (Eq. \ref{ims}).
The quasiparticle approximation can be improved by calculating the 
dispersive contribution  to the real part of the 
self-energy (second term  in Eq. (\ref{resi}))
\cite{eqp,rop,baldo,vo}.
Thus the single-particle energies are modified and the scheme must be 
iterated until the real part of the 
self-energy stabilizes. This is however still not a
 self-consistent scheme since the imaginary part of the 
self-energy is neglected in the 
calculation of the T-matrix.

\section{Self-consistent spectral function}

In this section we present the numerical results for the in-medium T-matrix
and the nucleon self-energy in the self-consistent calculation
 and in the quasiparticle
approximation.
 The calculations are performed
in a very simple separable rank one Yamaguchi potential \cite{yama}~:
\begin{equation}
V(p,p^{'})=\sum_{\alpha} \lambda_\alpha g_\alpha(p)g_\alpha(p^{'})
\end{equation}
 in the $^{1}S_{0}$ and 
$^{3}S_{1}$ waves, using the form factors
\begin{equation}
g_\alpha(p)=\frac{1}{p^2+\gamma^2} \ , \ \ \ \ \gamma=285.9 {\rm MeV}
\end{equation}
with the same strength $\lambda_\alpha$ of the attractive
potentials as used in \cite{alm}. 
This allows us to relate our results to alternative calculations using the 
quasi-particle approximation without using a realistic but complicated 
interaction.
We have used a kinematical limit in the
 momentum integrations, limiting the momentum of any nucleon to $|p|<1200$ MeV,
with the grid spacing of around $10$ MeV.
\begin{figure}
\begin{center}
\epsfig{file=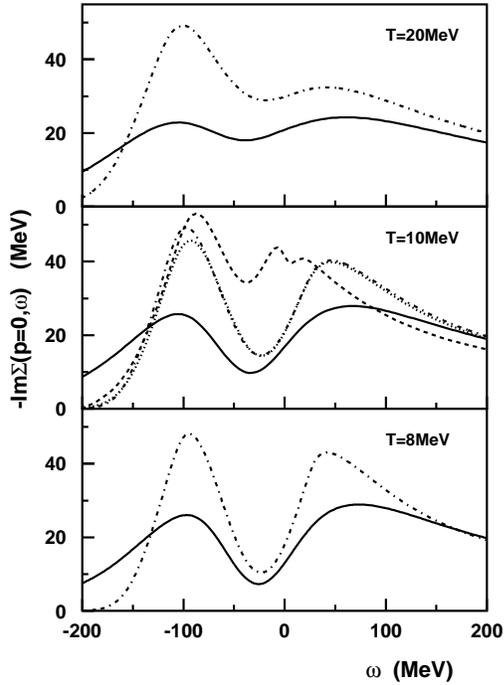,width=0.45\textwidth}
\end{center}
\caption{The imaginary part of the retarded 
self-energy as a function of  energy for $p=0$  at 
normal nuclear density and at different temperatures.
 The solid lines denote the self-consistent self-energies and the 
dashed-dotted lines denote the results of the quasiparticle approximation. 
For $T=10$ MeV the dotted line denotes the result 
when  taking the Hartree-Fock energy 
as the real part of the self-energy and a constant width of the spectral 
function of $6$ MeV.
The dashed line denotes the result of taking the self-consistent 
real part of the self-energy but a constant width of $6$ MeV in the
imaginary part of the self-energy.}
\label{g0fig}
\end{figure} \noindent
The effect of the cutoff in the free scattering changes the bounding energy
 of the deuteron to $2.1$ MeV.
The energy integrals were performed in the range $|\omega|<2400$ MeV with
the  smallest grid spacing  $1.75$ MeV at low temperature or 
for calculations using a preset small single-particle width.
In the present version of the calculation we assumed that for all
momenta
the width of the nucleon spectral functions is sufficiently large so that 
a direct integration in energies is possible. This limits the calculation
to nonzero temperatures, since as small temperatures the single-particle 
width approaches zero at the Fermi energy like $T^2$. Obviously, another 
limitation comes from the appearance of the pairing instability at
 small temperatures in the T-matrix scheme. In practice the lowest 
possible temperature (using only the $S$ wave interaction) that we could 
get stable iterations for is $8$ MeV. It is still significantly above 
 the critical temperature for the pairing transition, which is 
around $4$ MeV in the quasiparticle approximation with the assumed interaction
\cite{alm}. However the single-particle width obtained at $T=8$ MeV
 approaches the 
spacing grid at the limits of the kinematical region (due to the limitation
to  $S$ waves) and near the Fermi momentum (an expected property of an
 interacting Fermi gas).
All the in-medium calculation here presented have been performed at normal 
nuclear density.

It is easy to notice that 
the generalization of the T-matrix calculation to
include off-shell intermediate nucleon propagators does not change
qualitatively the fact that a pairing instability appears at some
critical temperature. The imaginary part of the T-matrix with off-shell 
propagators also vanishes at $\omega=2\mu$ and at some temperature a pole 
appears in the real part of the T-matrix. 
As mentioned above we could not study numerically the vicinity of the
 critical temperature 
in the self-consistent calculation.

In Fig. \ref{g0fig} the imaginary part of the self-energy at zero momentum 
is given for different temperatures as a function of  energy. In all the cases
 the self-consistent results are very different from the quasiparticle 
approximation. The self-consistent self-energies having generally a smaller
 imaginary part.
As the temperature is lowered more structure is visible  in the self-energy.
Beyond the minimum at the Fermi energy ($\mu \simeq -21$ MeV at $T=10$ MeV)
a second maximum appears around the energy $2\mu-\omega_0$, 
both in the self-consistent and in the quasiparticle approximation. 
In Fig. \ref{g0fig} for $T=20$ MeV
 the dashed line represents the imaginary part
of the self-energy obtained using the same real part of the self-energy 
as in the self-consistent solution but with the imaginary part of the 
self-energy fixed at $6$ MeV. The small width of the spectral function mimics 
in that case the $\delta$ function of the spectral function in the 
quasi-particle approximation. The resulting ${\rm Im}\Sigma^+$ 
is again very different from the self-consistent solution. It proves that
 it is not sufficient to take into account the correct real part of the 
self-energy  (the same as in the self-consistent solution) and neglect the
imaginary part of the self-energy. 
The full spectral function for the nucleon with momentum and frequency
dependent single-particle width must be taken for the intermediate propagators
 in the T-matrix equation (\ref{teq}).
For $T=20$ MeV we plot also the result obtained by taking only the 
Hartree-Fock self-energy for the real part of the self-energy, i.e. the same 
as in the quasiparticle approximation and a constant width in the spectral 
function of $6$ MeV. The result is very close to the 
one obtained in the quasiparticle approximation,
 meaning that a spectral function of width $6$ MeV can be approximated by 
a $\delta$ function. However, the true spectral function is different and
cannot be taken as a $\delta$ function.

\begin{figure}
\begin{center}
\epsfig{file=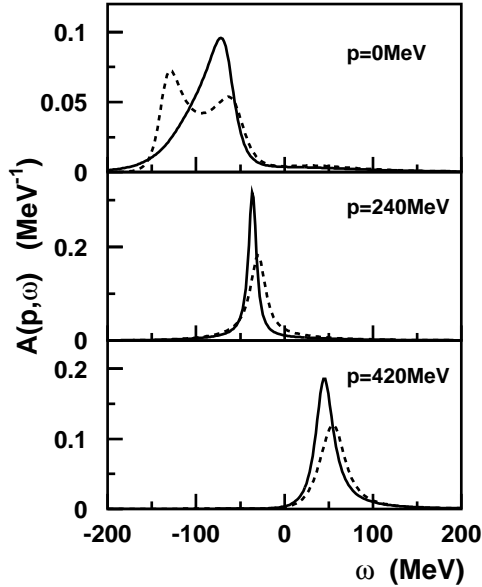,width=0.45\textwidth}
\end{center}
\caption{The nucleon spectral function as a function of energy 
at normal nuclear density and at 
$T=10$ MeV for three different momenta. The solid and the dashed 
lines denote the 
self-consistent and the quasiparticle approximations results respectively.}
\label{specfig}
\end{figure}

The differences in the imaginary part of the self-energy lead to significant 
differences in the spectral function. In Fig. \ref{specfig} the spectral 
functions are plotted for the quasiparticle and the self-consistent
 calculations. These results demonstrate that 
the nucleon spectral function cannot be calculated in the quasiparticle 
approximation. The position of the quasiparticle pick is different in the two 
calculation. It is due to differences in the real part of the self-energy.
The Hartree-Fock energies are slightly different because of different momentum 
distributions. Also the dispersive 
contribution to the real part of the self-energy are different because  
they originate from very different imaginary parts of the self-energies in the
two calculations. However, not only the positions of the picks in the spectral
functions are different, also their widths and shapes are different.  
Improvements on the calculation of the real part 
of the self-energy will not lead to a correct spectral function as long as 
the proper width of the intermediate propagators in the 
T-matrix equation is  not taken into account, giving the correct 
self-consistent real {\it and } imaginary parts of the self-energy.

\begin{figure}
\begin{center}
\epsfig{file=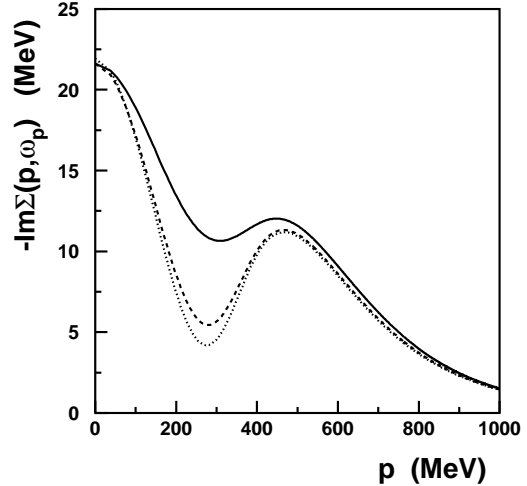,width=0.45\textwidth}
\end{center}
\caption{The imaginary part of the retarded 
self-energy at the quasiparticle pole as a function of  momentum at 
normal nuclear density. All the results are for
 the self-consistent calculation. The solid,
 dashed and dotted line represent ${\rm Im} \Sigma^+$ at $20$, $10$ and $8$ MeV
 respectively.}
\label{imselffig}
\end{figure}

\begin{figure}
\begin{center}
\epsfig{file=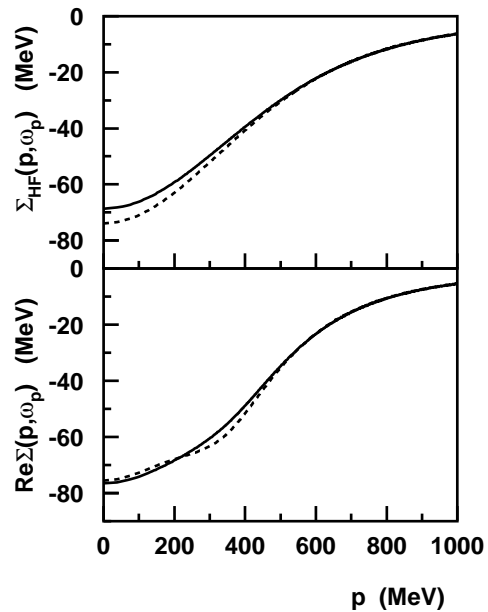,width=0.45\textwidth}
\end{center}
\caption{ The Hartree-Fock energy (upper panel) and
the real part of the retarded 
self-energy on shell (lower panel) as  functions of  momentum
at normal nuclear density and at temperatures of $20$ MeV (solid lines)
and $8$ MeV (dashed lines).}
\label{selffig}
\end{figure}

In Fig. \ref{imselffig} the imaginary part of the self-energy at the 
quasiparticle pole as a function of momentum is given. The decrease of the 
single-particle width with the temperature and a 
minimum around the Fermi energy can be observed.
In Fig. \ref{selffig} the Hartree-Fock energy and the complete real part of 
the self-energy are given at the quasiparticle pole as a function of 
momentum.
The behavior of the real part of the self-energy at the temperatures studied 
is relatively smooth.
 Only at the lowest temperature $T=8$ MeV a wiggle starts 
to appear similar to the one observed in \cite{alm}.
The momentum distributions are more diffuse for off-shell nucleons, leading
to smoother dependence of the Hartree-Fock energy on  momentum than in the
quasiparticle approximation \cite{alm}.

\begin{figure}
\begin{center}
\epsfig{file=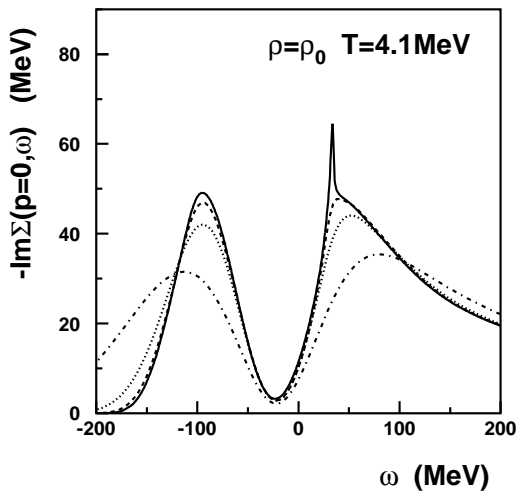,width=0.45\textwidth}
\end{center}
\caption{The imaginary part of the retarded 
self-energy as function of  energy for $p=0$  at 
normal nuclear density and $T=4.1$ MeV. The solid line denotes the
 quasiparticle approximation and the dashed, dotted and dashed-dotted lines
represent the results obtained using a fixed single-particle width of $6$,
 $15$ and $40$ MeV respectively.}
\label{t41fig}
\end{figure}

To close this section we would like to make one observation concerning the 
behavior of the self-energy around the pairing transition.
At the critical temperature of the pairing transition a pole appears in the 
T-matrix for pairs with zero total momentum both in the self-consistent and 
in the quasiparticle calculations. In the quasiparticle approximation it 
leads  to a singularity in the imaginary part of the self-energy \cite{alm}.
In the calculation of the self-energy using the off-shell propagators 
(\ref{ims}) there is one more energy integration which washes out this 
singularity. As an illustration we show in Fig. \ref{t41fig} the imaginary 
part of the self-energy at $T=4.1$ MeV in the quasi-particle approximation,
compared to results obtained using a fixed width of the spectral function
and the same Hartree-Fock energy as in the quasi-particle calculation. We
 observe that the singularity which starts to build up at $T=4.1$ MeV in 
the quasi-particle approximation is no longer present if finite widths of
propagators are taken. Note that it is the off-shellness of the propagator in
the calculation of the self-energy (\ref{ims}) not in the T-matrix equation,
which causes the singularity in ${\rm Im}\Sigma^+$ to disappear.

\section{In-medium cross sections}

The modeling of the nonequilibrium evolution in a heavy ion reaction by 
semiclassical transport models requires the knowledge of the in-medium cross 
section. For nucleons on-shell one can define the scattering cross sections
 similarly as in vacuum, but using an effective mass \cite{cs}.
If the nucleons are off-shell the definition of the cross section must be 
modified because the outgoing waves are localized in space \cite{di2}.
In this  section we will use a simplistic view of the scattering cross 
section as a parameter in the semiclassical collision integral.
The applicability of the quasiparticle 
approximations in the description of the nonequilibrium dynamics 
is questionable if the equilibrium calculations indicates the need for 
self-consistent off-shell calculations. However, 
in order for the complex dynamical evolution to be tractable we have to 
restrict ourselves to  quasiparticle transport models of heavy ion 
collisions. One can however take into account  in-medium modifications 
of the effective cross sections. In particular when calculating the
in-medium cross sections at equilibrium one can take 
into account the full propagators in the ladder diagrams.

The semiclassical collision term for a nucleon of momentum $p_1$
and energy $\omega_1$ has the form \cite{paw1}
\begin{eqnarray}
\int \frac{d \omega_2}{2 \pi}\frac{d^3 p_2}{(2 \pi)^3}
\int \frac{d \omega_3}{2 \pi}\frac{d^3 p_3}{(2 \pi)^3}
\int \frac{d \omega_4}{2 \pi}\frac{d^3 p_4}{(2 \pi)^3}
\nonumber \\
(2 \pi)^4 \delta^3({\bf p}_1+{\bf p}_2-{\bf p}_3-{\bf p}_4) 
\delta(\omega_1+\omega_2-\omega_3-\omega_4) \nonumber \\
\big|<{\bf k}|T^+({\bf P},\omega_1+\omega_2)|{\bf k}^{'}>_A \big|^2
 \nonumber \\ 
A({\bf p}_1,\omega_1) A({\bf p}_2,\omega_2) 
A({\bf p}_3,\omega_3) \nonumber \\
\{ \big(1-f({\bf p}_1,\omega_1)\big)\big(1-f({\bf p}_2,\omega_2)\big)
f({\bf p}_3,\omega_3)f({\bf p}_4,\omega_4) \nonumber \\ -
f({\bf p}_1,\omega_1)f({\bf p}_2,\omega_2)
\big(1-f({\bf p}_3,\omega_3)\big)\big(1-f({\bf p}_4,\omega_4)\big)
\} \ .
\end{eqnarray}
where ${\bf P}={\bf p}_1+{\bf p}_2$, ${\bf k}=({\bf p}_1 - {\bf p}_2)/2$,
${\bf k}^{'}=({\bf p}_3 - {\bf p}_4)/2$.
When putting the scattering particles on-shell and integrating over the 
angle\footnote{we use only the $S$ wave in the present work.}
 we can define an effective cross section $\sigma$ in the collision term
\begin{eqnarray}
\int \frac{d^3 p_2}{(2 \pi)^3}  \frac{|{\bf p}_1-{\bf p}_2|}
{M^*(P,k)}
\ \ \sigma \ \ \nonumber \\
 \{  \big(1-f({\bf p}_1)\big)\big(1-f({\bf p}_2)\big)
f({\bf p}_3)f({\bf p}_4) \nonumber \\  -
f({\bf p}_1)f({\bf p}_2)\big(1-f({\bf p}_3)\big)\big(1-f({\bf p}_4)\big)
\} \ ,
\end{eqnarray}
where
\begin{equation}
 M^*(P,k)= \frac{\partial\langle \omega_{p_1}+\omega_{p_2}\rangle_\Omega}
{\partial k}
\end{equation}
 is the effective mass. The sign $\langle \dots \rangle_\Omega$ denotes 
the averaging over the angle.
 The   cross section is
\begin{eqnarray}
\label{t2}
\sigma(P,k)=\frac{M^*(P,k)^2}
{4 \pi}
 \nonumber \\
\langle \ 
\big|<k|T^+(P,\omega_{p_1}+\omega_{p_2})
|k>_A \big|^2 \ \rangle_\Omega
\ .
\end{eqnarray}
\begin{figure}
\begin{center}
\epsfig{file=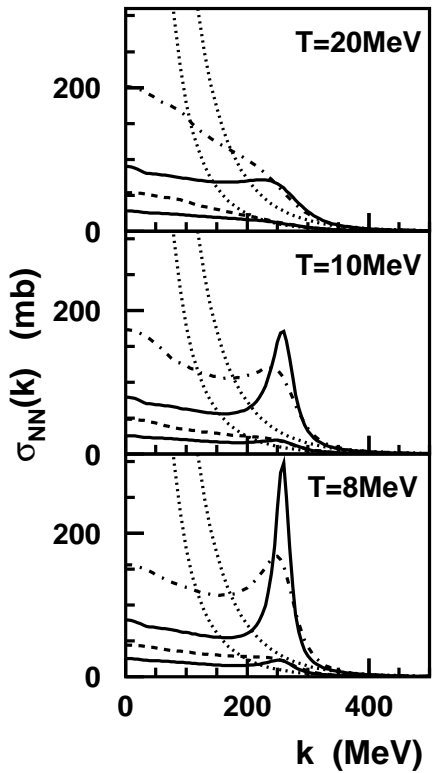,width=0.45\textwidth}
\end{center}
\caption{The in-medium cross sections for the scattering of
on-shell quasiparticles 
with zero total momentum as functions of the c.m. momentum at three different
 temperatures and at normal nuclear density. The dashed-dotted and the dashed 
lines denote the $n$-$p$ and the $n$-$n$ cross sections respectively 
calculated with the self-consistent in-medium T-matrix. The solid lines denote 
the corresponding in-medium cross sections calculated with the T-matrix in the
quasiparticle approximation and the dotted lines represent the corresponding
free cross sections.}
\label{csfig}
\end{figure} \noindent 
For the  T-matrix in the quasiparticle approximation
we can use the optical theorem to get
\begin{eqnarray}
\sigma(P,k)=-\frac{M^*(P,k)}{k}
 \nonumber \\
\frac{
<k|{\rm Im} T^+(P,\omega_{p_1}+\omega_{p_2})
|k>_A}{\langle 1 -f(\omega_{p_1})-f(\omega_{p_2})\rangle_\Omega}
\ .
\end{eqnarray}
However for the self-consistent T-matrix with off-shell propagators
the optical theorem is more complicated and  it is easier to use directly
Eq.  (\ref{t2}) to calculate the cross section.

In Fig. \ref{csfig} are presented the results for the $n$-$p$ and $n$-$n$
cross sections. The in-medium cross  sections calculated with the 
self-consistent T-matrix are generally larger than the cross sections obtained 
in the framework of the quasiparticle approximation.
Only part of the difference can be explained by a factor originating in the 
difference in the effective masses in the two calculations.
Both cross sections present a resonance-like pick in the $n$-$p$ scattering 
related to the pairing resonance (above $T_c$). These kind of 
structures in the energy dependence of the cross sections do not influence 
the evolution of the system, since the cross sections are always integrated
over momentum \cite{phcs}. However, the overall average cross sections in Fig. 
\ref{csfig}  are different
 and could lead to different transport properties in the collision. 
Before definite conclusions can be drawn calculations should be repeated with 
a realistic interaction.

\section{Conclusions}

We have presented a self-consistent calculation of the in-medium T-matrix.
The intermediate propagators in the T-matrix equation 
are full off-shell propagators. Both the real and imaginary part
of the self-energy in these propagators have been obtained consistently
from the T-matrix. The coupled system of equation was solved by iteration 
for the case of a simple separable interaction. The results were compared to 
a calculation using the quasiparticle approximation for the intermediate 
propagators in the T-matrix equation.

The imaginary parts of the self-energies and  the spectral 
functions obtained in the two calculations are very different. One cannot 
calculate reliably the spectral functions without using consistently the same 
spectral function through the whole approximation scheme. The width of the
 self-consistent spectral function is generally smaller than its 
quasiparticle estimates. The self-consistent T-matrix also presents 
a pairing instability 
at some critical temperature. However in the vicinity of the critical 
temperature the self-consistent iteration procedure cannot be performed
numerically. 

We have  calculated the corresponding 
in-medium cross sections in the
two approximations. The obtained cross sections are  different and
indicate that a self-consistent resummation of the ladder diagrams may be
important for a correct estimation of transport properties of the 
nuclear matter at finite temperature.

The spectral functions are very sensitive to  approximations on the 
imaginary part of the propagators in the T-matrix equation. In 
the present work we have demonstrated both the necessity and the feasibility of
a full self-consistent ladder diagram resummation when calculating the spectral
functions in the nuclear matter at finite temperature.
The method here presented could  also be applied to  Brueckner
 type calculations.  Before applying the iteration procedure to zero or 
low temperature, an explicit method of energy integration for
 quasiparticles around
the Fermi energy must be implemented. 
Similar methods must be used to overcome the limitations in the momenta of
 nucleons due to the cutoff.  Nucleons with large momenta are important
for the description of  short range correlations.
Since generally the width of  the spectral function is smaller at low 
temperature, one can expect that 
 the effect of the off-shell nucleon propagation would be 
less dramatic 
at zero temperature.

 \acknowledgments
The author is grateful to 
Pawe\l{} Danielewicz for enriching discussions.
He would also like to thank the NSCL for hospitality.
This work was partly supported by the National Science Foundation
under Grant PHY-9605207.

\vfill

\newpage

\newpage

\newpage

\newpage

\newpage

\newpage

\newpage

\end{document}